\documentclass[aps,pra,preprint,superscriptaddress]{revtex4-2}
\usepackage{amsmath,amssymb,graphicx,bm,hyperref,xcolor}
\usepackage[T1]{fontenc}
\newcommand{\D}[1]{\mathcal{D}[#1]}
\newcommand{\Tr}{\operatorname{Tr}}
\newcommand{\ket}[1]{\lvert #1\rangle}
\newcommand{\bra}[1]{\langle #1\rvert}
\renewcommand{\r}{\tau_+}
\renewcommand{\l}{\tau_-}

\begin{document}

\title{Directional pumping of a two-level system by a fluctuation-regulated quantum source}

\author{Sarfraj Fency}
\email{smjf21ip029@iiserkol.ac.in}
\author{Rangeet Bhattacharyya}
\email{rangeet@iiserkol.ac.in}
\affiliation{Department of Physical Sciences, Indian Institute of Science Education and Research Kolkata, Mohanpur 741246, India}

\begin{abstract}
A resonant classical field drives a two-level system coherently and, with ordinary
relaxation, produces saturation rather than directional population transfer. Here we show
that directionality emerges when the excitation source is retained as a quantum subsystem
and subsequently traced out. We model the source as a single-excitation degree of freedom,
or equivalently a stream of two-level ancillas, exchange-coupled to a target two-level
system. The source first moment generates the usual coherent drive, while its second
moments produce upward and downward dissipative channels. For a number-diagonal, fully
excited source, the first moment vanishes and only the upward channel remains, yielding
monotonic pumping of the target. The same structure follows from both an exact collision
map and a fluctuation-regulated second-order quantum master equation. For exponentially
decaying source correlations, the pumping rate shows a Lorentzian detuning dependence
governed by the correlation time $\tau_c$, accompanied by a dispersive frequency shift. In
the presence of thermal relaxation, the steady-state excited-state population can exceed
the resonantly driven optical-Bloch limit of one half. Within the fluctuation-regulated
master equation, this directional pumping arises as a drive-induced dissipative
contribution whose strength and spectral response are controlled by $\tau_c$.
\end{abstract}

\maketitle

\section{Introduction} \label{sec: Intro}
Optical pumping provides a paradigmatic example of directional population transfer induced by polarized electromagnetic radiation and is conventionally discussed in terms of multilevel atomic structure, polarization-selective excitation, and irreversible decay among magnetic or hyperfine sublevels \cite{kastler1950quelques, happer1967effective, happer1972optical}. In its familiar implementation, repeated cycles of selective excitation and spontaneous emission redistribute the populations until they accumulate in a state that is weakly coupled, or dark, to the applied polarization. Consider a system prepared in a ground sublevel with $m = -1/2$. A circularly
	polarized ($\sigma^+$) photon carries an energy $\hbar\omega_\circ$ and one unit of angular
	momentum $\hbar$ along the quantization axis. If the system has an excited sublevel separated from
	the initial state by exactly this energy and by exactly one unit of angular momentum, the photon
	is absorbed, since both the conservation laws are satisfied simultaneously.
	
	Once excited, the system may return to the state from which it came by re-emitting a photon of the
	same character. This is an elastic process and the internal state is left unchanged. On the other
	hand, through its coupling to the vacuum modes, the system may spontaneously emit a photon and
	decay to a different ground sublevel. This is an inelastic process and the internal angular
	momentum quantum number is permanently altered. We note that the system, once in the sublevel $m =
	+1/2$, can no longer be addressed by the drive, since the absorption of a further $\sigma^+$ photon
	would require an excited sublevel at $m = +3/2$ and no such state exists in the manifold. The state
	is therefore dark because the selection rules of the
	driving field fail to reach it. Each passage through the excited state therefore constitutes a
	stochastic branching with an absorbing boundary, and after several such cycles the population
	accumulates in the dark
	sublevel with near-unit efficiency. The steady state so obtained is, in the ideal case, a pure
	state. The entropy initially carried by the internal degrees of freedom is
	carried away by the spontaneously emitted photons. As such, the internal multilevel structure provides a natural route by which excitation and decay can address different transitions, and the resulting population transfer acquires an effective directionality.

	Historically, the subject has been built on exactly this premise. In his founding proposal,
	Kastler noted that an unequal population of the magnetic sublevels may be produced optically,
		provided the ground state atoms are paramagnetic, i.e. $J \neq 0$ or $F \neq 0$
		\cite{kastler1950some}. The condition for pumping is therefore stated in the first paper of the
		field as a condition on the internal structure. Confirmations followed shortly, in the double
		resonance experiments of Brossel and Bitter \cite{brossel1952new} and in the ground state pumping
		reported by Brossel, Kastler, and Winter \cite{brossel1952greation}. The technique subsequently
		developed through the works of Hawkins \cite{hawkins1953polarization}, Dehmelt
		\cite{dehmelt1957slow, dehmelt1958spin}, Bell and Bloom \cite{bell1957optical}, and Franzen and
		Emslie \cite{franzen1957atomic}. The modern theoretical treatment originates in the work of Barrat
		and Cohen-Tannoudji, who recast the problem in the density matrix formalism and showed that under
		optical excitation the ground state acquires a finite lifetime and a self-energy
		\cite{barrat1961etude, barrat1961elargissement}. The reviews and the monographs on the subject
		\cite{happer1972optical,cohen1966optical,happer2010optically,auzinsh2010optically,cohen1992atom}
		are organized around the hyperfine and the Zeeman manifolds, the branching ratios, and the
		selection rules, and the same structure underlies the techniques that followed, such as the
		sub-Doppler
		laser cooling \cite{dalibard1989laser}, the spin-exchange optical pumping
		\cite{walker1997spin,gentile2017optically}, the optical magnetometry \cite{budker2007optical}, and
		the vapour-cell atomic clocks \cite{gozzelino2023realization}.

A strict two-level system (TLS), however, does not possess an internal branching structure. When it is driven by a resonant classical field, the rotating-wave Hamiltonian may be written as $H_{\rm cl}= \Omega (\sigma_+ + \sigma_-)/2,$
and the two directions of stimulated exchange enter with equal amplitude. With spontaneous relaxation rate $\Gamma$, the optical Bloch equations give (see Appendix \ref{sec: OBE})
$\rho_{ee}^{\rm ss}=\Omega^2/(\Gamma^2+2\Omega^2)\leq 1/2,$
and, consequently, a stationary inversion cannot be generated by the coherent field alone. The classical field can at best saturate the transition, at which point the two levels are equally
populated and the steady state is maximally mixed, i.e. the state of infinite temperature. Instead
of extracting the entropy from the system, the field has maximized it. The reason appears
structural. With only two levels, the field and the dissipation connect the same pair of states,
and no state can remain inaccessible to the field. The mismatch between the field and the
dissipation that
 the pumping requires would then be possible only when the internal Hilbert space is labelled by
more than one quantum number, i.e. the energy and the angular momentum. We note that this familiar bound is specifically a statement about a \emph{classically prescribed coherent drive} together with ordinary relaxation; it does not prohibit a TLS from being pumped by a nonequilibrium quantum reservoir.
	
The restriction above is also tied to the usual description of the drive. In the standard treatment, the drive enters as a classical $c$-number field, for example as $\Omega\cos(\omega t)\sigma_x$. The same first-order Hamiltonian contains both $\sigma_+$ and $\sigma_-$ with equal amplitudes and is therefore reciprocal: it drives absorption and stimulated emission with equal strength. A coherent drive of this form cannot by itself generate a preferred direction of population transfer.

In this work, we treat the source that supplies the drive energy as a quantum degree of freedom and trace it out only after its interaction with the target. To this end, we model the source as a quantum system coupled to the two-level system through
a Jaynes-Cummings interaction of the form $g(\l\sigma_+ + \r\sigma_-)$. Written in this way,
the coupling has two distinct channels. A quantum removed from the source raises the system, and a
quantum returned to the source lowers it, and these two processes are weighted by different matrix
elements of the source operators. We then take trace over its degrees of freedom and retain the calculation up to the
second order in $g$, thereby obtaining a dissipator which is structurally similar to the
$T_1$ and $T_2$ terms arising from the usual system-environment coupling, but which originates from
the source. This viewpoint is closely connected with repeated-interaction models, micromaser physics, and cascaded quantum systems, where quantum emitters or propagating quantum fields act as dynamical sources rather than as prescribed $c$-number amplitudes \cite{jaynes1963comparison, filipowicz1986theory, ciccarello2022quantum, gardiner1993driving, carmichael1993quantum, baragiola2012n}. 

A phenomenological pumping model would simply contain the incoherent channels $\sigma_+$ and $\sigma_-$ with unequal rates; no first-order coherent drive is required. The steady state is then determined by the ratio of the two rates. In the present treatment these channels are not introduced phenomenologically, but are obtained from the quantum source after tracing over its degrees of freedom. We use the fluctuation-regulated quantum master equation (FRQME) to obtain this second-order contribution. Within the FRQME, the applied interaction generates drive-induced dissipation and the finite correlation time regulates its strength and gives the associated dispersive response \cite{frqme,didexpt,chanda2020optimal}. In the present problem this drive-induced dissipation is the pumping channel, and the fluctuation timescale directly regulates the rate of directional population transfer.

The source state enters the reduced dynamics through its first and second moments. A source having a nonzero transition expectation value produces the usual coherent Hamiltonian drive at first order, whereas a number-diagonal excited source has a vanishing first moment but unequal second moments. We show that these second moments distinguish the ability of the source to donate an excitation from its ability to accept one and, upon tracing over the source, generate unequal upward and downward dissipative rates. In particular, for a fully excited source the downward channel vanishes, and the target is consequently pumped monotonically toward its excited state. Thus, the state of the quantum source fixes the direction of transfer, while the FRQME regulator fixes the strength of the resulting second-order pumping process.

The source polarization and the direction of energy transfer play different roles. A fixed circular polarization, or helicity $q=\pm1$, selects the spherical component of the target transition to which the source couples; however, for a strict TLS there is only one addressed transition, and reversing the helicity does not by itself interchange upward and downward pumping. The \emph{direction} of transfer is controlled by the source occupation, while the polarization determines which angular-momentum transition is accessible. This distinction is not merely semantic and becomes particularly important when the construction is extended to a multilevel target.

The manuscript is organized as follows. In Sec. \ref{sec: Model}, we present the model. In Sec.
\ref{sec: frqme}, we give a brief description of the FRQME and use it to obtain the steady state solution. We present the numerical results in Sec. \ref{sec: results}. We then discuss the physical content of the
mechanism and compare it with the literature in Sec. \ref{sec: discussion}, and finally conclude in Sec. \ref{sec: conclusion}.

\section{Quantum-source model} \label{sec: Model}
\subsection{Target and fixed-helicity source}
To construct the minimal model, the target TLS is described by
\begin{equation}
H_S=\frac{\omega_0}{2}\sigma_z, \qquad \sigma_+=\ket e\bra g, \qquad \sigma_-=\ket g\bra e.
\end{equation}
For a selected helicity $q$, the minimal two-level source Hilbert space is taken as
\begin{equation}
\mathcal H_C=\mathrm{span}\{\ket0,\ket{1_q}\},
\end{equation}
where $\ket0$ denotes no available excitation and $\ket{1_q}$ one excitation in the selected channel. Defining
\begin{equation}
\tau_-=\ket0\bra{1_q},\qquad \tau_+=\ket{1_q}\bra0,
\end{equation}
the exchange interaction in the rotating-wave approximation is
\begin{equation}
V= g\left(\sigma_+\otimes\tau_-+\sigma_-\otimes\tau_+\right).
\label{eq:exchange}
\end{equation}
Equation~\eqref{eq:exchange} is an analogue of the Jaynes--Cummings interaction \cite{jaynes1963comparison}. We note, however, that a literal bosonic mode differs from this source in an important respect: an occupied bosonic mode can accept an additional excitation, whereas $\tau_+\ket{1_q}=0$. As such, the strict one-way map derived below should be understood as describing a source, a stream of single-excitation ancillas, or an effective cascaded emitter, and not an unrestricted single bosonic mode.

\subsection{Source state and its moments}
In the ordered basis $\{\ket0,\ket{1_q}\}$, write the source state as
\begin{equation}
\rho_C=\begin{pmatrix}1-p & c^*\\ c & p\end{pmatrix},
\qquad 0\le p\le1,
\qquad |c|^2\le p(1-p).
\label{eq:source_state}
\end{equation}
Here, the population $p$ denotes the probability that an incoming source unit carries an excitation, while the corresponding source coherence is
\begin{equation}
\langle\tau_-\rangle=c,
\end{equation}
while the relevant second moments are
\begin{equation}
\langle\tau_+\tau_-\rangle=p,
\qquad
\langle\tau_-\tau_+\rangle=1-p.
\label{eq:moments}
\end{equation}
For a pure source state one may write
\begin{equation}
\ket{\psi_C}=\sqrt{1-p}\,\ket0+e^{i\phi}\sqrt p\,\ket{1_q},
\end{equation}
so that $|c|=\sqrt{p(1-p)}$. The first-order target Hamiltonian obtained by tracing Eq.~\eqref{eq:exchange} over the source is therefore
\begin{equation}
H_{\rm coh} = g\left(c\sigma_+ +c^*\sigma_-\right).
\label{eq:firstmoment}
\end{equation}
For a fully excited source, $p=1$ and necessarily $c=0$. It can therefore supply energy without providing a coherent phase reference, and its leading action on the target is dissipative rather than a conventional Rabi drive.

\subsection{Exact one-collision benchmark}
Before invoking a coarse-grained master equation, the directional character of the interaction can be established directly from a single collision. To this end, let one excited ancilla interact resonantly with the target for a time $\delta t$, with $\vartheta=g\delta t$. In the one-excitation sector,
\begin{equation}
\ket{g,1_q}\longrightarrow \cos\vartheta\,\ket{g,1_q}-i\sin\vartheta\,\ket{e,0},
\end{equation}
whereas $\ket{e,1_q}$ cannot stimulate an additional source excitation because $\tau_+\ket{1_q}=0$. Thus, after tracing over the ancilla, the reduced map of the target is
\begin{equation}
\Phi(\rho)=K_1\rho K_1^\dagger+K_0\rho K_0^\dagger,
\end{equation}
with
\begin{equation}
K_1=\ket e\bra e+\cos\vartheta\ket g\bra g,
\qquad
K_0=-i\sin\vartheta\,\sigma_+.
\end{equation}
For $\vartheta\ll1$,
\begin{equation}
\Phi(\rho)-\rho=\vartheta^2\D{\sigma_+}(\rho) + O(\vartheta^4),
\label{eq:collisionweak}
\end{equation}
where $\mathcal{D}[L](\rho) = L\rho L^\dagger - \frac{1}{2}L^\dagger L \rho - \frac{1}{2}\rho L^\dagger L$ is the Lindblad dissipator.

Consequently, a stream of freshly prepared ancillas arriving at a rate $r$ produces an effective pumping rate $\Gamma_{\rm coll}=r\sin^2\vartheta\simeq r g^2\delta t^2$. This provides the standard repeated-interaction route from a microscopic excitation-exchange process to a Markovian dissipator \cite{filipowicz1986theory,ciccarello2022quantum}; moreover, it makes explicit that a sustained pumping process necessarily requires either a reset of the source or a continuous flux of fresh excitations.

\section{Fluctuation-regulated reduced dynamics} \label{sec: frqme}
\subsection{Regulated second-order source term}
To connect the above microscopic picture with drive-induced dissipation, the source contribution is now treated within the FRQME. In the original construction, stochastic fluctuations of the local environment, together with ensemble averaging and coarse graining, generate an exponential regulator for the second-order terms \cite{frqme}. Following the same structure for the present quantum source, the reduced equation may be written as

\begin{eqnarray}\label{eq:frqme_source}
	\dot\rho_s(t)&=&- i \, [H_S+H_{\rm coh}, \, \rho_s]
	\\
	&-&\int_0^\infty d \tau \,R(\tau)\,
	\Tr_C\left[V(t),\left[V(t-\tau),\rho_s(t)\otimes\rho_C\right]\right],
	\nonumber
\end{eqnarray}
where $R(\tau)$ is the correlation regulator. For an exponential envelope,
\begin{equation}
R(\tau)=e^{-\tau/\tau_c},\qquad \tau\ge0,
\end{equation}
and with detuning $\Delta \omega=\omega_0-\omega_C$,
\begin{equation}
\int_0^\infty d\tau\,e^{-\tau/\tau_c}e^{i\Delta \omega \, \tau} =\frac{\tau_c}{1-i\Delta \omega \, \tau_c} =J(\Delta \omega)+iS(\Delta \omega),
\end{equation}
where
\begin{equation}
J(\Delta \omega)=\frac{\tau_c}{1+\Delta \omega^2\tau_c^2},
\qquad
S(\Delta \omega)=\frac{\Delta \omega\tau_c^2}{1+\Delta \omega^2\tau_c^2}.
\label{eq:JS}
\end{equation}

The real part of the regulated response gives the dissipative contribution, while
the imaginary part gives the associated dispersive shift. The two therefore arise from the
same source response, as in the FRQME treatment of drive-induced damping and dynamic
shifts \cite{frqme,chanda2020optimal}.

Using the source moments in Eq.~\eqref{eq:moments}, and retaining the secular second-order contribution, one obtains
\begin{equation}
\mathcal L_C \rho_s = \kappa(\Delta \omega) \Big(p \, \D{\sigma_+}(\rho_s) +(1-p) \, \D{\sigma_-}(\rho_s) \Big),
\label{eq:DID}
\end{equation}
and
\begin{equation}
\kappa(\Delta \omega) = 2g^2J(\Delta \omega) = \frac{2g^2\tau_S}{1+\Delta \omega^2\tau_S^2},
\label{eq:kappa}
\end{equation}
where, $\tau_S$ is the correlation time associated with the source. The same integral generates a dispersive Hamiltonian correction with characteristic scale
\begin{equation}
\delta\omega_{\rm DID}(\Delta \omega) \sim g^2 \, S(\Delta \omega) = \frac{g^2 \Delta \omega \tau_S^2}{1+\Delta \omega^2 \, \tau_S^2},
\label{eq:shift}
\end{equation}
with the overall sign and the precise operator prefactor fixed by the interaction-picture convention. On resonance the dispersive contribution vanishes, and the population dynamics used below is unaffected by this convention.

Equation~\eqref{eq:DID} is the drive-induced dissipative term generated by the source interaction. The unequal source moments determine the relative strengths of the upward and downward channels, while Eq.~\eqref{eq:kappa} gives their fluctuation-regulated rate. On resonance, $\kappa(0)=2g^2\tau_S$ within the weak-coupling regime. At finite detuning, $\kappa(\Delta \omega)$ increases with $\tau_S$ for $|\Delta\omega|\tau_S<1$, is maximum at $|\Delta\omega|\tau_S=1$, and decreases for longer correlation times. Thus, $\tau_S$ controls both the pumping rate and its spectral width. Equation~\eqref{eq:DID} constitutes the central structural result of the present analysis. Here, the upward and downward rates are not introduced as independent phenomenological parameters; rather, they are inherited directly from the unequal second moments of the quantum source. For a fully excited source, $p=1$, only $\D{\sigma_+}(\rho_s)$ survives, whereas for a source in its ground state, $p=0$, only $\D{\sigma_-}(\rho_s)$ remains. At $p=1/2$ the source-induced dissipative contribution is unbiased, although it is not zero. The case $p=1/2$ should not be identified with a classical-field limit, since a classical coherent field corresponds instead to a different source state, typically a large-amplitude bosonic coherent state.

\subsection{Ordinary target relaxation}
To determine how the source-induced process competes with ordinary relaxation of the target, the thermal contribution is included as
\begin{equation}
\mathcal L_L\rho_s=\gamma_\downarrow \, \D{\sigma_-}(\rho_s) + \gamma_\uparrow \, \D{\sigma_+}(\rho_s),
\end{equation}
where detailed balance gives $\gamma_\uparrow/\gamma_\downarrow=e^{-\beta\hbar\omega_0}$. In the parametrization used for the numerical data,
\begin{equation}
\gamma_\downarrow=\Gamma_L P_g,
\qquad
\gamma_\uparrow=\Gamma_L P_e,
\qquad
P_{g,e}=\frac{e^{\pm\beta\hbar\omega_0/2}}{2\cosh(\beta\hbar\omega_0/2)},
\end{equation}
with $\Gamma_L=\omega_{\text{SL}}^2\tau_L$ and $\tau_L$ being the correlation time associated with the local environment. The complete resonant generator is then
\begin{equation}
\dot\rho_s= - i \, [H_{\rm coh},\rho_s] + \Gamma_\uparrow \, \D{\sigma_+}(\rho_s) + \Gamma_\downarrow \, \D{\sigma_-}(\rho_s),
\label{eq:master}
\end{equation}
where
\begin{equation}
\Gamma_\uparrow=\kappa(\Delta \omega) \, p+\gamma_\uparrow,
\qquad
\Gamma_\downarrow=\kappa(\Delta \omega) \, (1-p)+\gamma_\downarrow.
\label{eq:rates}
\end{equation}

\subsection{Steady state and pumping threshold}
For the steady-state analysis, the source phase is chosen as the transverse reference, so that Eq.~\eqref{eq:firstmoment} can be written as a resonant drive with the Rabi frequency
\begin{equation}
\Omega=2g|c|.
\end{equation}
Let $\Gamma_\Sigma=\Gamma_\uparrow+\Gamma_\downarrow$. The exact stationary excited-state population of Eq.~\eqref{eq:master} is
\begin{equation}
\rho_{ee}^{\rm ss}=\frac{\Gamma_\uparrow\Gamma_\Sigma+\Omega^2}
{\Gamma_\Sigma^2+2\Omega^2},
\label{eq:rhoess}
\end{equation}
and the stationary coherence has magnitude
\begin{equation}
|\rho_{eg}^{\rm ss}|=
\frac{\Omega\,|\Gamma_\downarrow-\Gamma_\uparrow|}
{\Gamma_\Sigma^2+2\Omega^2}.
\label{eq:cohss}
\end{equation}
The phase of $\rho_{eg}^{\rm ss}$ is fixed by the phase of $c$.

In the fully excited-source limit, $p=1$ and $c=0$, and therefore
\begin{equation}
\rho_{ee}^{\rm ss}=\frac{\kappa+\gamma_\uparrow}{\kappa+\gamma_\uparrow+\gamma_\downarrow}.
\label{eq:p1ss}
\end{equation}
gives complete population transfer. When the thermal channel is retained, the population exceeds the resonantly driven optical-Bloch value of one half provided
\begin{equation}
\kappa>\gamma_\downarrow-\gamma_\uparrow.
\label{eq:threshold_general}
\end{equation}
This is an inversion threshold, i.e. the condition for $\rho_{ee}^{\rm ss}>1/2$; the source-induced pumping channel itself is present for any nonzero $\kappa$.
On resonance, using $\kappa=2g^2\tau_S$ and $\Gamma_L=\omega_{\text{SL}}^2\tau_L$, this becomes
\begin{equation}
2g^2\tau_S>
\omega_{\text{SL}}^2\tau_L\tanh\!\left(\frac{\beta\hbar\omega_0}{2}\right).
\label{eq:threshold}
\end{equation}
For the equal-correlation-time parametrization $\tau_S = \tau_L = \tau_c$, Eq.~\eqref{eq:threshold} reduces directly to the threshold used in the numerical calculations presented below and Eq. ~\eqref{eq:p1ss} reduces to
\begin{eqnarray}
	\rho_{ee}^{\rm ss}= \frac{2 g^2 + \omega_{\text{SL}}^2 P_e}{2 g^2 + \omega_{\text{SL}}^2}
	\label{eq:p1ss_reduced}
\end{eqnarray}

\subsection{First-moment coherence and second-moment bias}
For a pure source, the coherence and population are not independent, since $|c|=\sqrt{p(1-p)}$. To make this relation explicit, define the source population bias
\begin{equation}
A=2p-1.
\end{equation}
Then
\begin{equation}
2|c|=\sqrt{1-A^2}.
\label{eq:tradeoff}
\end{equation}
Equation~\eqref{eq:tradeoff} therefore gives an exact kinematic tradeoff within this family of source states: the maximum occupation bias is obtained for a number state, for which the coherent first moment vanishes, whereas the largest possible first moment occurs when the occupation bias is zero. We note that this relation should not be interpreted as establishing an equivalence between a one-quantum ancilla and a classical coherent field; rather, it shows how the same two-state source continuously redistributes its action between a coherent first-order contribution and a biased second-order contribution.

\begin{figure*}[t]
	\centering
	\includegraphics[width=0.9\textwidth]{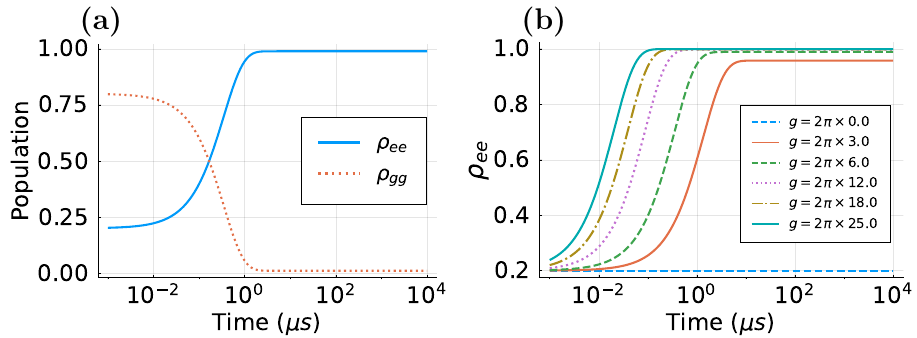}
	\caption{Directional population transfer for a fully excited source. (a) The excited-state population rises monotonically while the ground-state population is depleted. (b) Increasing the effective source-target coupling accelerates the transfer and raises the stationary population against a fixed thermal relaxation channel.}
	\label{fig:time}
\end{figure*}

\section{Numerical results} \label{sec: results}
The analytical results are now illustrated using the numerical calculations supplied with the manuscript, which were generated from the same effective master-equation structure. Unless stated otherwise, the parameters are $\omega_0=2\pi\times5$ G rad s$^{-1}$, $g=2\pi\times6$ M rad s$^{-1}$, $\omega_{\text{SL}}=2\pi\times1$ M rad s$^{-1}$, $T=37$ mK, and an effective correlation time $\tau_c=10^{-9}$ s.

Figure~\ref{fig:time}(a) shows the characteristic dynamical signature of the $p=1$ limit. The calculation starts from the thermal equilibrium population. Since $c=0$ in this limit, the first-order coherent drive vanishes, while the source-induced upward channel in Eq.~\eqref{eq:DID} remains finite; as a result, the population evolves irreversibly toward the excited state and saturates. The transfer is monotonic and shows no nutation, in contrast to the damped oscillation obtained from the optical Bloch equations. This is the expected behavior of an irreversible rate process.

Fig.~\ref{fig:time}(b) shows the effect of increasing $g$ on the pumping efficiency and the characteristic pumping time. When $g = 0$, there is no pumping. As $g$ is increased, the population is pumped to the excited state, although not completely, since the environmental channel continues to compete. Once the condition in Eq.~(\ref{eq:threshold}) is crossed, the excited-state population exceeds one half. It approaches unity only when the source-induced pumping rate dominates the ordinary relaxation rate. The rate at which the pumping proceeds depends on the coupling strength, and the characteristic pumping time decreases as $g$ is increased, in agreement with the on-resonance weak-coupling scaling $\kappa(\Delta \omega)\propto g^2\tau_c$. We define the characteristic pumping time as the first instance of time after which there is no appreciable change in the population of that particular atomic state.

\begin{figure*}[t]
	\centering
	\includegraphics[width=0.9\textwidth]{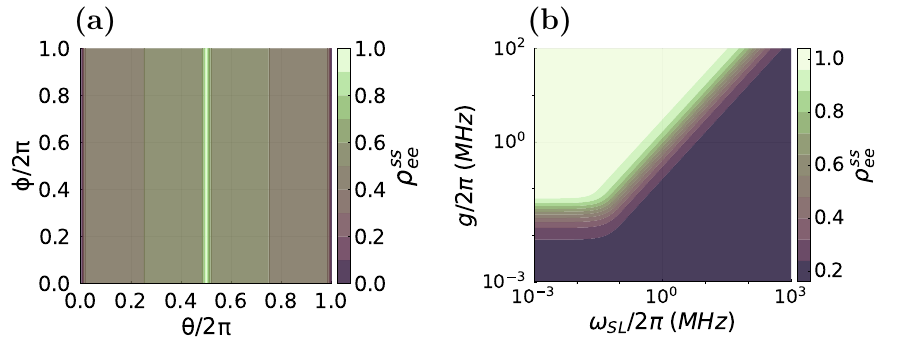}
	\caption{Steady-state population and dynamical regimes. (a) $\rho_{ee}^{\rm ss}$ as a function of the two angles used to parametrize a pure source state. The narrow bright ridge at $\theta=\pi$ corresponds to $p=1$, while the result is independent of the source phase $\phi$ in this fully incoherent limit. (b) Saturated excited-state population versus effective source-target coupling and system-environment coupling. The diagonal crossover is the numerical manifestation of Eq.~\eqref{eq:threshold}.}
	\label{fig:phase}
\end{figure*}

For the pure-state parametrization $p=\sin^2(\theta/2)$, the fully excited source corresponds to $\theta=\pi$, while the phase $\phi$ changes the coherent first moment without changing the source populations. Fig.~\ref{fig:phase}(a) displays a phase-independent ridge of maximal pumping at $p=1$. Away from this ridge, a nonzero first moment coexists with both second-order channels and the stationary state is, in general, mixed. The independence of $\phi$ is expected, since $\phi$ sets the phase of the coherent drive and not the weight of either of the dissipative channels.

Figure~\ref{fig:phase}(b) shows the competition between the two physical processes: thermal relaxation, and source-induced pumping Three distinct regimes are visible in the contour. (i) For $\omega_{\text{SL}} \gg g$, the Eq. (\ref{eq:p1ss_reduced}) gives $\rho_{ee}^{\text{ss}} \approx 1-P_g = 0.2$ for the chosen parameters, which is the thermal value. (ii) For $\omega_{\text{SL}} \sim g$, the two physical processes compete and we obtain $\rho_{ee}^{\text{ss}} \approx 1 - P_g/3 = 0.73$. (iii) For $\omega_{\text{SL}} \ll g$, the source-induced pumping dominates and we obtain	$\rho_{ee}^{\text{ss}} \approx 1$. The slope of this crossover follows directly from the rate-balance condition in Eq.~\eqref{eq:threshold}.

\begin{figure*}[t]
	\centering
	\includegraphics[width=\textwidth]{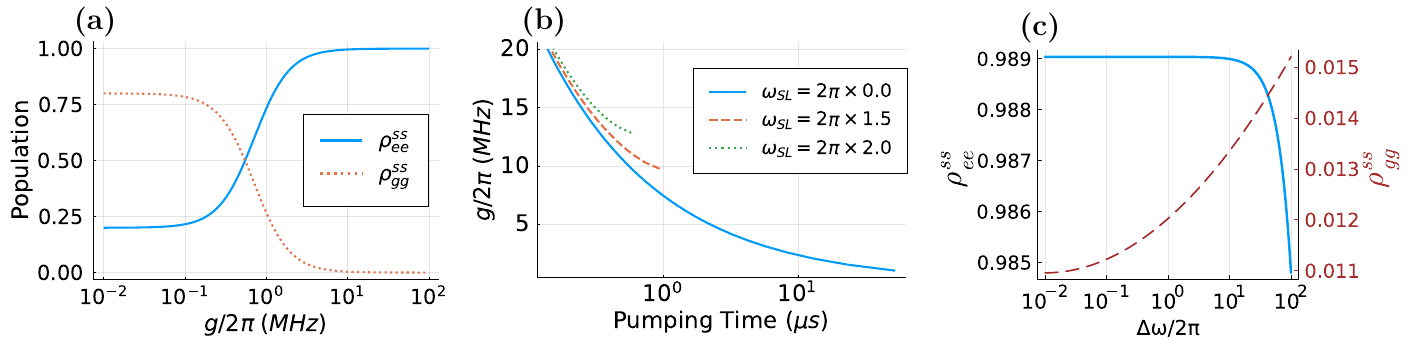}
	\caption{Coupling, pumping time, and detuning dependence. (a) Steady-state populations versus source-target coupling. (b) Pumping time versus coupling for several environmental relaxation strengths. (c) Steady-state populations versus detuning. The weak change near resonance is consistent with the Lorentzian factor in Eq.~\eqref{eq:kappa}: detuning primarily reduces the source-induced rate and therefore becomes important once $|\Delta \omega|\tau_c\gtrsim1$ or when the thermal channel is comparable to the source-induced one.}
	\label{fig:detuning}
\end{figure*}

In Fig.~\ref{fig:time}(b), the excited-state population increases with $g$ and crosses the inversion threshold, while the characteristic pumping time also changes.We quantify this by plotting the saturated value of the population as a function of $g$ in the Fig.~\ref{fig:detuning}(a). It is seen that the population crosses the inversion threshold and then approaches unity as $g$ is increased further.

Further, to see how the characteristic pumping time changes as we change the coupling strength, we allowed the system to evolve and saturate for a particular value of $g$. The time at which the excited state population was $\geqslant 0.99$ is plotted in the Fig.~\ref{fig:detuning}(b). It is observed that the characteristic pumping time decreases as the coupling strength is increased, and that it increases with $\omega_{\text{SL}}$, since the environmental channel opposes the transfer.

Figure~\ref{fig:detuning}(c) shows the detuning dependence of the fluctuation-regulated pumping process. The steady-state population varies only weakly near resonance, as follows directly from Eq.~\eqref{eq:rates}. Both source-induced rates contain the same Lorentzian factor $J(\Delta \omega)$ and, therefore, detuning does not directly modify their ratio; for a nearly fully excited source, however, it weakens the desired upward channel relative to the independent thermal relaxation channel. The effect becomes appreciable only when $\Delta \omega \tau_c \gtrsim 1$. The robustness of the pumped state to the detuning is therefore a prediction of the mechanism and not an accident of the chosen parameters.

\begin{figure*}[t]
	\centering
	\includegraphics[width=\textwidth]{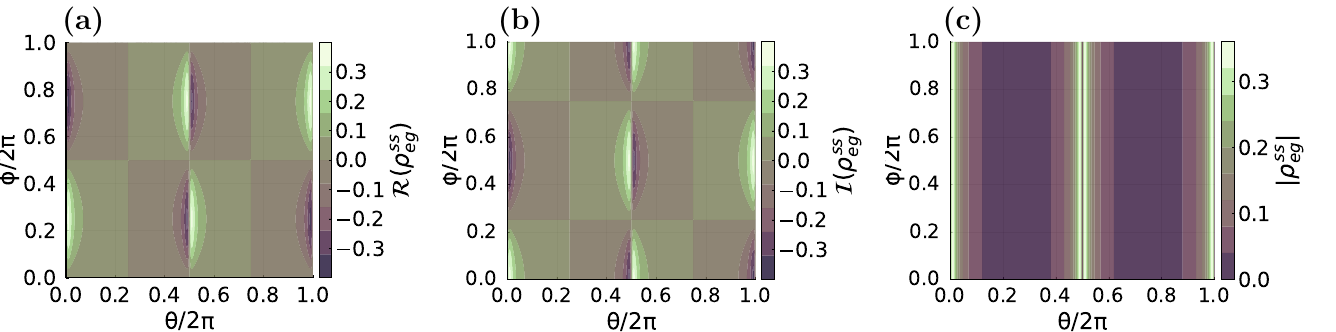}
	\caption{Steady-state target coherence as a function of the pure-source-state angles. Panels (a) and (b) show the real and imaginary parts of $\rho_{eg}^{\rm ss}$ and panel (c) its magnitude. The coherence vanishes at the population eigenstates of the source, including the fully excited pumping point $\theta=\pi$, in agreement with Eqs.~\eqref{eq:cohss} and \eqref{eq:tradeoff}.}
	\label{fig:coherence}
\end{figure*}
	
The characteristic width of the pumping response is controlled by the source correlation time, while Eq.~\eqref{eq:shift} predicts a dispersive frequency shift across the same resonance. The same analytical expression also shows how changing $\tau_c$ changes the population transfer. At resonance, increasing $\tau_c$ enhances the second-order pumping rate as $2g^2\tau_c$, as long as the weak-coupling and coarse-graining conditions remain valid. Away from resonance, the dependence is non-monotonic because the numerator and the Lorentzian denominator in Eq.~\eqref{eq:kappa} are both controlled by $\tau_c$. The dissipative and dispersive responses originate from the real and imaginary parts of the same regulated integral, and a simultaneous measurement of both quantities under a controlled variation of the source linewidth would therefore provide a substantially stronger test of the present picture than population transfer alone. 

Figure~\ref{fig:coherence} shows the steady-state coherence of the target. At a source population eigenstate, where $c=0$, the target acquires no steady-state coherence even though a finite second-order pumping channel remains; moving away from this point introduces a phase-sensitive first-order term and, correspondingly, a residual target coherence. At the fully directional point the source first moment vanishes and no steady-state target coherence survives. Away from this point, however, a directional population bias and a residual coherence can coexist. Further, this affords a degree of control, since a partial pumping with a prescribed residual coherence may be obtained by a suitable choice of the state of the source. We note that this numerical behavior follows the analytical distinction between the coherent first moment and the number-diagonal second moments of the source, and therefore provides a direct visualization of the crossover between a phase-coherent drive and a directional pump source. 

\section{Discussion} \label{sec: discussion}
The upward Lindblad channel $\D{\sigma_+}(\rho_s)$ is standard in quantum optics and reservoir engineering \cite{scully1997quantum,poyatos1996quantum,diehl2008quantum,verstraete2009quantum}, and directional transfer from nonequilibrium quantum sources is also known from repeated-interaction and cascaded-system descriptions \cite{filipowicz1986theory,gardiner1993driving,carmichael1993quantum,ciccarello2022quantum}. Here, the source-induced channel is obtained from the second-order drive term of the FRQME, and its strength is regulated by the fluctuation timescale. Keeping the source explicit also allows the coherent and dissipative contributions to be related to different moments of the same source state.

The first moment $c$ generates the coherent nutation of the target, while the second moments $p$ and $1-p$ determine the upward and downward dissipators. A number-diagonal source can transfer population without producing a conventional Rabi term, and the transition from coherent driving to directional pumping is encoded directly in the state of the source rather than being imposed through two unrelated terms in the target master equation.

The present formulation separates the role of polarization from that
of source inversion. A selected helicity labels the angular-momentum channel in
Eq.~\eqref{eq:exchange} and may allow or forbid a particular Zeeman transition, depending
on the chosen pair of magnetic sublevels and the propagation convention; it does not, by
itself, make absorption irreversible. The one-way character instead originates from the
source occupation asymmetry.

The present treatment has three main points. First, the pumping term is derived from the source interaction and its coefficient is fixed by $g$, the source state $p$, and the fluctuation correlation time $\tau_c$. Vanier earlier used a Redfield treatment of random perturbations with a chosen correlation time to describe optical pumping as a second-order dissipative process \cite{vanier1969optical}; here the second-order term is generated by the applied source interaction and appears as drive-induced dissipation in the FRQME. Second, the pumping rate depends explicitly on $\tau_c$, which can in principle be tested by changing the source linewidth, in the same spirit as the measurement of drive-induced dissipation \cite{didexpt}. Third, the same source state determines both the coherent and dissipative terms: its first moment produces nutation, while its second moments determine the directional pumping. The coherent limit therefore cannot pump a strict TLS, whereas the fully excited number-diagonal limit gives a purely dissipative pump.

The FRQME introduces a finite correlation-time regulator which makes the dissipative rate and the dispersive shift the real and imaginary parts of the same response function, as given by Eqs.~\eqref{eq:kappa} and \eqref{eq:shift}. The correlation time $\tau_c$ therefore acquires a direct physical significance: it determines the source-induced linewidth, controls the on-resonance pumping rate within the weak-coupling regime, and fixes the scale of the associated dispersive response. In the short-correlation-time limit, $\kappa(\Delta \omega)\rightarrow0$ and the second-order source-induced pumping is suppressed; the remaining steady state is then fixed by the first-order coherent drive, when present, and by ordinary relaxation. We also note that the formal scaling $\kappa(0)\propto\tau_c$ cannot be extrapolated to $\tau_c\rightarrow\infty$, because once $g \, \tau_c$ is no longer small the Markov and coarse-graining assumptions cease to remain valid, and coherent source-target exchange is expected to re-emerge.

The two-level source assumption is physically significant rather than merely technical. If the source is replaced by a bosonic mode initially containing one photon, a target already in $\ket e$ can undergo stimulated emission into a two-photon state, and the exact one-way collision map no longer follows. A literal propagating photon source should therefore be treated within a cascaded-system or Fock-state input-output description \cite{gardiner1993driving,carmichael1993quantum,baragiola2012n}. To this end, the present source TLS is most naturally interpreted as an elementary emitter or a resettable single-excitation ancilla whose output couples to the target; moreover, this interpretation makes explicit the need for a source flux in order to sustain the pumping process.

The role of source fluctuations in the polarization degree of freedom provides a further extension of the present picture. A helicity-dependent dephasing process can suppress coherence between the $q=+1$ and $q=-1$ polarization modes and, as such, select the circular basis as a pointer basis; however, it cannot by itself remove an unwanted helicity population. Pure helicity therefore still requires appropriate preparation or reset, polarization-dependent loss, or a target selection rule in addition to dephasing. It would be particularly interesting to construct a microscopic stochastic source Hamiltonian in which the same fluctuations both suppress the polarization coherence and generate the correlation envelope appearing in Eq.~\eqref{eq:frqme_source}, since such a construction would place the polarization selection and the fluctuation-regulated target dynamics within a common microscopic description.
The same construction may also be useful in other driven open systems in which second-order drive terms compete with ordinary relaxation.

Several limitations of the present treatment should be mentioned. We have taken the correlation timescales of the source and of the	local environment to be equal; the steady state for unequal timescales is given in Appendix	\ref{sec: unequal tau}, although a complete treatment would require a derivation of the FRQME with	two independent fluctuation processes, which we leave to a future work. The underlying mechanism	is not tied to the two-level	realization considered here and can be systematically extended to multilevel systems. Such an	extension would allow the selection rules and the source inversion to act together, with the	polarization determining which transition is addressed and the inversion determining the direction	of the transfer. We expect this to be the physically relevant configuration, and we leave it to a	future work.

The stationary master equation assumes weak coupling, a separation between the correlation and relaxation time scales, and a continuously refreshed source, while the numerical calculations use an effective source coupling and an exponential regulator rather than a fully propagated source field. A quantitatively complete optical implementation would therefore require a benchmark against a cascaded emitter or a Fock-state wave-packet calculation, together with an independent characterization of the source flux, linewidth, and polarization. Such a comparison would be particularly useful if the independently measured source parameters could be used to predict both the pumping rate and the associated dispersive shift without introducing a fitted phenomenological pump rate.

\section{Conclusion} \label{sec: conclusion}
In this work, we have shown that a two-level target can undergo directional population pumping when the excitation source is treated as a nonequilibrium quantum subsystem rather than as a prescribed coherent field. Within the minimal two-level system model, the first moment of the source produces the coherent Hamiltonian drive, whereas its second moments determine the strengths of the upward and downward dissipative channels; as such, a fully excited and number-diagonal source has no coherent first moment and generates a purely upward channel in the ideal limit. We have further shown that ordinary thermal relaxation competes with this process, which leads to the threshold in Eq.~\eqref{eq:threshold} for exceeding the resonantly driven optical-Bloch population bound.

Within the FRQME, the source-induced channel is governed by a finite correlation time. The Lorentzian pumping profile and its dispersive partner arise from the same regulated response and are therefore tied directly to the correlation properties of the source. The central result is therefore that directional pumping is generated here by the drive-induced dissipative term of the FRQME, while the fluctuation correlation time controls its rate, linewidth, and competition with ordinary relaxation.

\begin{acknowledgments}
S.F. acknowledges financial assistance from the University Grants Commission, Government of India, through a Junior Research Fellowship (reference ID BORSU00637050 A).
\end{acknowledgments}

\bibliographystyle{apsrev4-2}
\bibliography{manuscript}

\appendix
\section{Optical-Bloch bound for a resonantly driven TLS}\label{sec: OBE}

Consider a TLS with ground state $\vert g\rangle$ and excited state $\vert e\rangle$, driven on resonance. In the rotating frame and after the rotating-wave approximation, the drive Hamiltonian is
\begin{equation} \label{OBE_H}
	H_{\rm cl} = \frac{\Omega}{2}\,(\sigma_+ + \sigma_-).
\end{equation}

With spontaneous-emission rate $\Gamma$, the density matrix obeys
\begin{equation} \label{OBE_ME}
	\dot{\rho} = -i[H_{\rm cl},\rho] + \Gamma\,\mathcal{D}[\sigma_{-}](\rho),
\end{equation}

At steady state,
\begin{align}
	0&=-\Gamma\rho_{ee}-\frac{i\Omega}{2}(\rho_{ge}-\rho_{eg}),\\
	0&=-\frac{\Gamma}{2}\rho_{eg}-\frac{i\Omega}{2}(1-2\rho_{ee}).
\end{align}
Eliminating $\rho_{eg}$ gives
\begin{equation}
	\rho_{ee}^{\rm ss} = \frac{\Omega^2}{\Gamma^2+2\Omega^2} \leq \frac12 \quad \text{for all } \Omega
\end{equation}
with $\rho_{ee}^{\text{ss}}\rightarrow1/2$ only as $\Omega/\Gamma\rightarrow\infty$. For $\Gamma=0$ there is no relaxation to a unique steady state; any stationary state under the resonant Hamiltonian has equal level populations. The value $1/2$ is therefore not a general upper bound for a TLS coupled to a nonequilibrium reservoir.

\section{Trace over the source at second order}
In the interaction picture,
\begin{equation}
V(t)= g\left(\sigma_+\tau_-e^{i\Delta \omega t}+\sigma_-\tau_+e^{-i\Delta \omega t}\right).
\end{equation}
Substituting this interaction into the double commutator of Eq.~\eqref{eq:frqme_source}, and subsequently tracing over the source, produces terms proportional to the two source correlations
\begin{align}
C_+(s)&=\Tr_C\!\left[\tau_+\tau_-\rho_C\right]e^{i\Delta \omega s}=p\,e^{i\Delta \omega s},\\
C_-(s)&=\Tr_C\!\left[\tau_-\tau_+\rho_C\right]e^{-i\Delta \omega s}=(1-p)e^{-i\Delta \omega s}.
\end{align}
After the secular approximation, the real parts yield
\begin{eqnarray}
\mathcal L_C\rho_s&=&\,2g^2J(\Delta \omega)p\left(\sigma_+\rho_s\sigma_- -\frac12\{\sigma_-\sigma_+,\rho_s\}\right)\\
										&+&2g^2J(\Delta \omega)(1-p)\left(\sigma_-\rho_s\sigma_+ -\frac12\{\sigma_+\sigma_-,\rho_s\}\right), \nonumber
\end{eqnarray}
which reproduces Eq.~\eqref{eq:DID}. Moreover, the imaginary parts of the same regulated integrals generate the corresponding dispersive second-order Hamiltonian correction.

\section{Unequal source and environmental correlation times} \label{sec: unequal tau}
In the numerical simulations, the correlation timescales of the drive source and of the local environment have been taken to be equal. Here we record the steady state when the two are different. Let $\tau_S$ denote the correlation time associated with the source and $\tau_L$ that associated with the local environment. It is convenient to define the two total rates,
\begin{equation}
\kappa(\Delta \omega)=\frac{2g^2\tau_S}{1+\Delta \omega^2\tau_S^2},
\qquad
\Gamma_L=\omega_{\text{SL}}^2\tau_L.
\end{equation}
Equations~\eqref{eq:rates}--\eqref{eq:rhoess} remain valid with these rates. In particular, for $p=1$ and $c=0$,
\begin{equation}
\rho_{ee}^{\rm ss}=\frac{\kappa(\Delta \omega)+\Gamma_LP_e}{\kappa(\Delta \omega)+\Gamma_L}.
\end{equation}
At resonance, the inversion criterion reduces to Eq.~\eqref{eq:threshold}, while the equal-time parametrization used in the numerical calculations is recovered by setting $\tau_S=\tau_L=\tau_c$.  We note that the principal relations of this work do not depend on the two timescales being equal. The pumping asymmetry
	\begin{equation}
		\frac{\Gamma_{\uparrow} -	\Gamma_{\downarrow}}{\kappa(\Delta\omega)}= 2p-1
	\end{equation}
is a ratio taken within the drive sector	alone, in the absence of the environmental relaxation, and hence $\kappa(\Delta\omega)$ cancels from it for any $\tau_S$. Similarly, $\vert c	\vert = \sqrt{p(1-p)}$ is a property of $\rho_C$ alone. Further, for a fully inverted source with	$\Gamma = 0$, the Eq.~\eqref{eq:rhoess} gives $\rho_{ee}^{\text{ss}} = 1$ irrespective of $\tau_S$, and hence the purity of the fixed point is also unaffected.
	
The threshold, however, is modified as given in Eq.~\eqref{eq:threshold}. The ratio $\tau_S/\tau_L$ enters as an additional parameter. On the logarithmic axes of the Fig.~\ref{fig:phase}(b), the boundary separating the first and the third regimes retains its slope but is displaced rigidly by $\frac{1}{2}\ln(\tau_L/\tau_S)$. Moreover, a degeneracy present in the main text is lifted, since for $\tau_S = \tau_L$ the steady state depends only on the ratio	$g/\omega_{\text{SL}}$, whereas the two rates $\kappa(\Delta \omega)$ and $\Gamma_L$ are independently tunable in the general case.

Finally, the detuning enters only through $\kappa(\Delta \omega)$, and hence the half-width of the response shown in the Fig. \ref{fig:detuning}(c) is set by the source alone,
	\begin{equation} \label{halfwidth}
		\Delta\omega_{1/2} = \frac{1}{\tau_S} \sqrt{\frac{2 g^2\tau_S}{\omega_{\text{SL}}^2 \tau_L \tanh\left(\beta\hbar \omega_{\circ}/2\right)} - 1}.
	\end{equation}
The detuning dependence of the pumped state can therefore provide a direct estimate of $\tau_S$.

We note that the treatment of this appendix assumes that the exponential regulator appearing in the drive-induced dissipator is governed by $\tau_S$ alone. In the FRQME, however, the regulator on all the second order terms originates from the construction of the coarse-grained propagator, in which the memory kernel of the environment is acquired irrespective of the Hamiltonian appearing in the second order \cite{frqme}. If the source is also allowed to fluctuate and is	traced over on the same footing, both the kernels may be expected to appear in the drive-drive	dissipator, and the effective regulator would then be shorter than either. A derivation of the	FRQME with two independent fluctuation processes is required to settle this point, and it is left	to a future work. Accordingly, the treatment with separate $\tau_S$ and $\tau_L$ should be regarded as an effective extension of the single-regulator FRQME used in the main text.

\end{document}